\documentclass[aps,prl,reprint,showpacs,superscriptaddress,floatfix]{revtex4-2}

\usepackage{CJK}
\usepackage{etoolbox}
\usepackage{graphicx,amssymb,bm,amsfonts,amsmath,graphics,color,epstopdf,float,wrapfig}
\usepackage[bookmarks=false,pdfstartview=FitH,colorlinks=true,citecolor=blue,linkcolor=blue]{hyperref}

\usepackage[version=3]{mhchem}
\usepackage{siunitx}

\begin{document}


\title{Coherent Interactions Between Silicon-Vacancy Centers in Diamond}
\author{Matthew W.\ Day}
\affiliation{Department of Physics, University of Michigan, Ann Arbor, MI 48109, USA}
\author{Kelsey M.\ Bates}
\affiliation{Department of Physics, University of Michigan, Ann Arbor, MI 48109, USA}
\author{Christopher L.\ Smallwood}
\affiliation{Department of Physics, San Jos\'e State University, San Jose, CA 95192, USA}
\affiliation{Department of Physics, University of Michigan, Ann Arbor, MI 48109, USA}
\author{Rachel C. \ Owen}
\affiliation{Department of Physics, University of Michigan, Ann Arbor, MI 48109, USA}
\author{Tim Schr{\"o}der}
\affiliation{Department of Physics, Humboldt-Universit{\"a}t zu Berlin, Newtonstra{\ss}e 15, 12489 Berlin, Germany}
\author{Edward Bielejec}
\affiliation{Sandia National Laboratories, Albuquerque, NM 87185, USA}

\author{Ronald Ulbricht}
\affiliation{Max Plank Institut f\"ur Polymerforschung, Ackermannweg 10, 55128 Mainz, Germany}
\author{Steven T.\ Cundiff}
\email[Email: ]{cundiff@umich.edu}
\affiliation{Department of Physics, University of Michigan, Ann Arbor, MI 48109, USA}
\date {\today}

\begin{abstract}
 We report coherent interactions within an ensemble of silicon-vacancy color centers in diamond. The interactions are ascribed to resonant dipole-dipole coupling. Further, we demonstrate control over resonant center pairs by using a driving optical pulse to induce collective, interaction-enabled Rabi-oscillations in the ensemble. Non-resonant center pairs do not undergo collective oscillations.
\end{abstract}

\maketitle


Color centers in diamond host localized and optically accessible electronic states. Some, such as the nitrogen-vacancy center and a range of Group-IV vacancies, feature favorable spin properties that attract wide interest as a platform for precision measurements \cite{Degen2017}, nanoscale sensors \cite{Sushkov2014,Nguyen2018}, and  building blocks for various components of quantum information infrastructure \cite{Sipahigil2016,Sukachev2017,Wan2020,Bhaskar2020}. The negatively charged silicon-vacancy center in diamond (SiV$^{-}$) is one such defect under intense study. Due to the similarity of the ground and excited state wavefunctions, SiV$^{-}$s are largely protected from the first-order perturbations that plague the more heavily studied nitrogen-vacancy center \cite{Dietrich2014,Liu2020,Londero2018,Meesala2018,Udvarhelyi2019}, and exhibit a sharp zero-phonon line (ZPL) into which a majority of the excited-state photoluminescence (PL) is concentrated. As such, SiV$^{-}$ centers are suitable candidates in which to generate and manipulate coherent interactions between discrete quantum systems.

The ability to harness interactions between color centers is a tantalizing prospect. It could enable multi-center quantum sensing \cite{Degen2017} to surpass the standard shot-noise limit imposed on measurements of \textit{N} independent qubits. Additionally, one proposed quantum information processing scheme relying on dipole-dipole interaction generated entanglement \cite{campbell2020} could be implemented in ensembles of color centers. The most common current approach to manipulating interactions between SiV$^{-}$s is to find two proximate centers with compatible transition frequencies and etch a diamond waveguide between them \cite{Evans2018}. This approach is promising, but entails large manufacturing overhead and could  be biased by the selection of exceptional color centers with favorable optical properties, giving an incomplete picture of the underlying interaction dynamics.

In this Letter we demonstrate that resonant dipole-dipole interactions between SiV$^{-}$s can coherently couple centers distinct centers. We investigate these interactions using a nonlinear double-quantum, two-dimensional (DQ2D) coherent spectroscopic probe of a high density ($\sim 10^{18} \text{cm}^{-3}$) ensemble of SiV$^-$s. DQ2D probes an ensemble nonlinear response thereby reducing potential single center selection bias and requiring no sample-specific tailoring to study interactions. DQ2D is also background-free, in the sense that there is no DQ2D signal in the absence of excitation dependent interactions \cite{Kim2009}. We show that these interactions are strong enough to cause the observed inhomogeneous linewidths, contributing to the overall optical response at dominant order. Further, we demonstrate control of inter-center interactions to cause coherent Rabi oscillations of center-center pairs by varying the strength of a pump pulse preceding the DQ2D probe pulses. These results show that dipole-dipole interactions can be used to generate and manipulate few-body quantum states in an ensemble of SiV$^-$ color centers. 

\begin{figure}
\includegraphics[width = .4\textwidth]{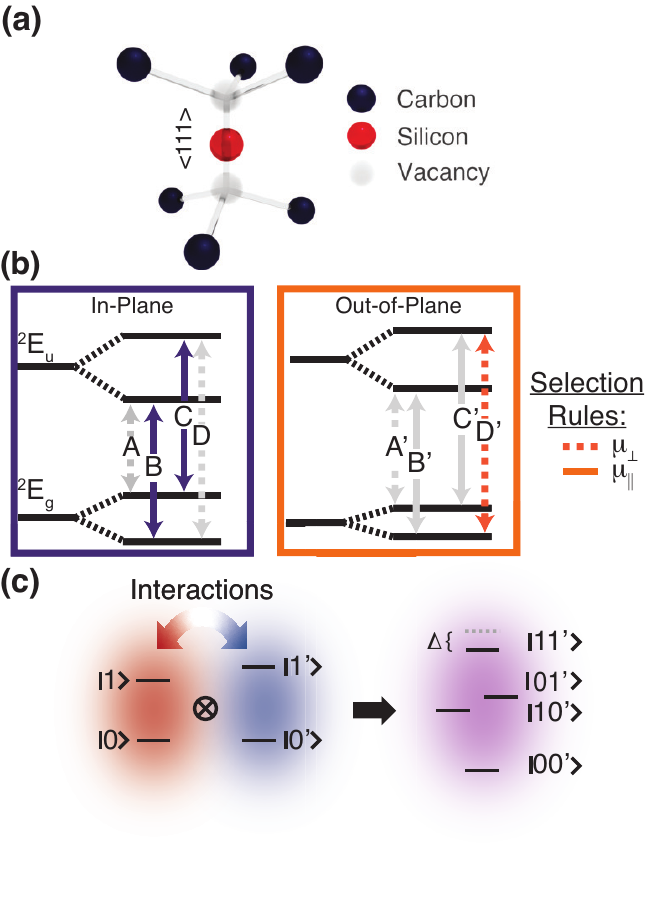}
\caption{\label{fig1}(a) The SiV system: a silicon interstitial between two vacancies. (b) The electronic level scheme with the ZPL transitions labelled. Polarization selection rules are depicted relative to the SiV$^-$ axis, grey arrows show missing transitions. (c) A depiction of how excitation-dependent interactions couple two systems.}
\end{figure}

Figure \ref{fig1} is a summary of the color center system. Figure \ref{fig1}(a) depicts the SiV$^-$ center with the vacancy-Silicon-vacancy axes orientation falling along the $\langle111\rangle$ direction in the lattice. The SiV$^-$ zero-phonon line (ZPL) transitions probed with DQ2D are marked B, C, and D' in Fig.~\ref{fig1}(b). In this sample, due to strain and geometry considerations, the polarization selection rules yield PL from only these three transitions for X-polarized excitation (horizontal with respect to the experimental axes) \cite{Bates2020}. Transitions B and C yield PL collected from centers oriented in the plane of the sample whereas transition D' yields PL originating from centers oriented out of the plane of the sample, as the averaged bulk strain shifts the peak locations for different center orientations relative to each other. For completeness, a DQ2D spectrum taken for the orthogonal excitation pulse polarization is included in the supplementary information \cite{supplement}.

In general, quasi-resonant excitation dependent interactions between non-identical two-level systems (in the presence of inhomogeneity) can be treated as a shift in the doubly-excited state of the joint system, as shown in Fig.~\ref{fig1}(c), leaving the singly-excited eigenstates unmodified. We let $\Delta = \Delta_d - i\Delta_s$, phenomenologically capturing both the interaction modified dephasing and energy shift of the doubly-excited state, commonly known as the excitation-induced dephasing (EID) and excitation-induced shift (EIS).
%
Our DQ2D method employs four pulses derived from a titanium sapphire oscillator (center wavelength 735 nm, repetition rate 75.5 MHz, pulse width 200 fs full-width, half maximum). Pulses are individually frequency tagged with acousto-optic modulators to allow the selection of multiple linear or nonlinear signal pathways. The four pulses illuminate the sample co-linearly, generating nonlinear polarizations and modulated populations in the SiV$^{-}$ electronic states \cite{Smallwood2016,Tekavec2007,Nardin2013,Smallwood2020}. The sample is tilted at a 30$^{\circ}$ angle to the horizontal and the modulated populations result in a modulated PL signal, collected in reflection \cite{Smallwood2020}. The four-wave mixing (FWM) pathway is selected by phase-synchronous detection of the fourth-order PL modulation at $\omega_{sig} =  \omega_1 +\omega_2 -\omega_3 - \omega_4$. 

\begin{figure}[h!]
\includegraphics[width = .4\textwidth]{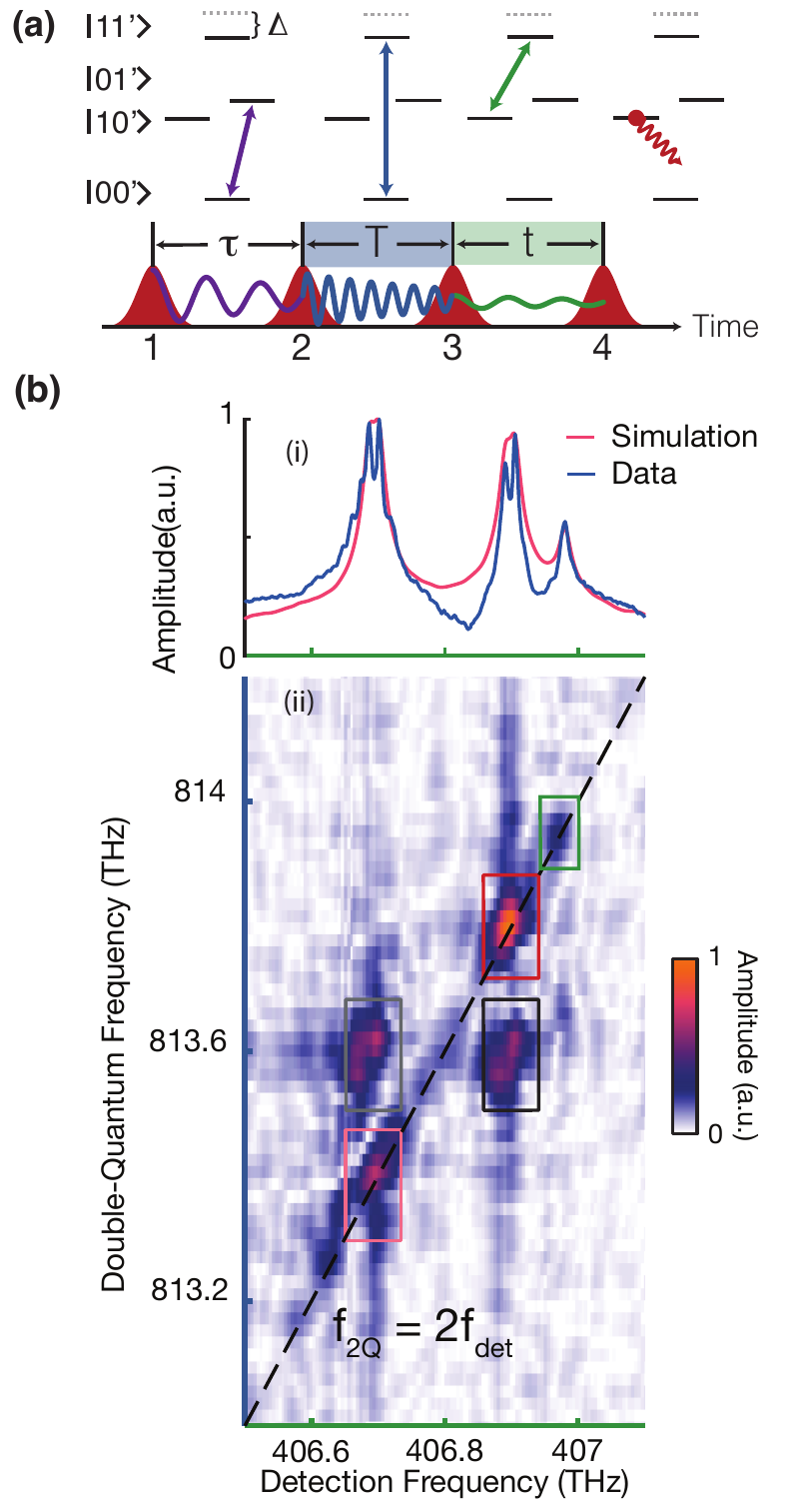}
\caption{(a) A diagram of the pulse sequence used in the DQ2D experiment. Coherences between center pairs evolve during waiting time $T$ and their excitation-induced influence on the PL spectrum is resolved along waiting time $t$. (b)(i) A comparison between a linear PL spectrum and a simulation incorporating dipole-dipole interactions as the broadening mechanism. (ii) A DQ2D spectrum from the SiV$^-$ ensemble. Peaks in the pink, red, and green boxes arise from coherent coupling between two resonantly excited states. Non-resonant coherent coupling between centers yields peaks on either side of the diagonal, in the black and grey boxes.}
\label{fig2}
\end{figure}

Figure \ref{fig2}(a) is a schematic of a single DQ2D signal pathway generated by this pulse sequence. In this case, the first pulse drives a coherence between the ground and first excited state, which is converted by the second pulse into a ``double-quantum" coherence between the ground and doubly-excited state at twice the energy. For this reason, the frequency-evolution along time delay \textit{T} is termed the ``Double-Quantum'' evolution of the FWM signal. The third and fourth pulses convert that coherence into a population in a singly (or doubly) excited state. The phase and amplitude of the PL modulation at $\omega_{FWM} $ are monitored as a function of \textit{T} and \textit{t}, then Fourier-transformed yielding two-dimensional spectra. The time delays corresponding to the Fourier-transform of the two time axes are color-coded in Fig.~\ref{fig2}(a) and the axes colors in Fig.~\ref{fig2} (b)(i) and (ii). 

Figure \ref{fig2}(b)(i) is a linear, PL detected absorption spectrum of the ZPL transitions. Figure \ref{fig2}(b)(ii) is a DQ2D spectrum showing interactions between both resonant and non-resonant centers, as evidenced by peaks along the diagonal and off-diagonal directions respectively. As mentioned above, these peaks are a background-free signature of excitation-dependent interactions between centers. Having demonstrated that color centers interact, we now turn to determining the potential interaction mechanism.

The most likely mechanisms by which SiV$^{-}$ centers interact are wavefunction overlap (Dexter coupling) or dipole-dipole mediated interactions. In the first case, we might expect an additional contribution to the DQ2D spectrum from the hybridized states \cite{specht2015}.  We rule out this possibility, because the resulting spatially weak hybridization has not been seen to contribute to DQ2D spectra \cite{Nardin2014}, and \textit{ab initio} simulations of the electron wavefunctions in SiV$^-$s do not show much spatial extent outside of roughly one unit cell, an order of magnitude less than required for significant spatial overlap between wavefunctions of adjacent centers at our implantation densities \cite{Udvarhelyi2019,Gali2013}. We therefore conclude that the most likely interaction mechanism is dipole-dipole coupling between adjacent centers.

Even in in the complex SiV$^{-}$ level system, signal pathways in DQ2D spectra will be dominated by separate two-level systems interacting. A detailed argument for this assertion can be found in the supplementary information \cite{supplement}, but the conclusion is that the DQ2D lineshape for a given pair of interacting states, following the conventions in Ref. \cite{Smallwood2016}, the state labelling scheme in Fig.~\ref{fig1}(c), and assuming approximately delta-function pulses, is 
\begin{equation}
\begin{split}
    S^{(3)}(\tau,\omega_T,\omega_t)  = \frac{\mu_{10}^2 \mu_{1'0'}^2}{8\hbar^3}\frac{e^{-i\Omega_{10'}\tau} + e^{-i\Omega_{01'}\tau}}{\omega_T -\Omega_{11'}-\Delta} & \\  \times \Big( \frac{1}{\omega_t - \Omega_{01'}} -\frac{1}{\omega_t - \Omega_{01'}- \Delta} + & \\ \frac{1}{\omega_t - \Omega_{10'}}  -\frac{1}{\omega_t - \Omega_{10'} - \Delta} \Big) &
\end{split}
\label{eq1}
\end{equation}
where $\Omega_{ij} = \omega_{ij} - i \gamma_{ij}$ is the complex frequency associated with the transition from the $ij$-th state to the ground state including phenomenological dephasing $ \gamma_{ij}$, $\omega_{ij} = (E_i - E_j)/\hbar$ is the transition center frequency, and $\mu_{ij}$ is the dipole moment corresponding to the transition between the $i$ and $j$ states, and $\Delta = \Delta_d - i\Delta_s$ is the complex interaction parameter, as defined above. Included in equation \ref{eq1} is the assumption that the singly and doubly excited states have the same transition dipole moments. If this is not the case, it will only apply an overall scaling to equation \ref{eq1}. From equation \ref{eq1}, it is clear that the nonlinear signal exactly cancels when $\Delta = 0$, in the case of non-interacting two-level systems, consistent with the claim that this is a background-free spectroscopic probe of interactions.

Though a useful probe of interactions, the DQ2D lineshape is highly sensitive to the interplay between inhomogeneous broadening and the average interaction strength between centers. This sensitivity can be seen by the dependence of the nonlinear DQ2D signal on $\omega_{ij}$, $\gamma_{ij}$ and $\Delta = \Delta_s - i \Delta_d$. In the case that interactions between states reduce dephasing (known as excitation-induced narrowing), $\gamma_{ij}$ (or $1/T_2^*$) and EID have opposite signs.

We remove this lineshape ambiguity by simply considering the linear PL detected absorption spectrum. We know from previous studies on this sample that the linear PL linewidths are dominated by inhomogenous broadening \cite{Smallwood2020}, so we monitor the PL modulated at a frequency $\omega_{sig} = \omega_{3} - \omega_{4}$ while varying time delay \textit{t} between the third and fourth pulses in our experiment. This yields a coherent linear spectrum corresponding to a PL-detected absorption measurement, presented in Fig.~\ref{fig2} (b)(i). Since the transition dipole moment is known \cite{Becker2017}, and since the only tuning parameter in the interaction simulation is the density (where we make the rough assumption that 10\% of the implanted ions yield color centers after annealing), we simulate the lineshape of the linear PL by generating a random distribution of centers reflective of the implantation density profile of our sample and diagonalizing the Hamiltonian
\begin{equation}
    H_{tot} = H_{0} + \sum_{i,j,i\neq j}H_{int,ij}
\end{equation}
where $H_0$ is the bare Hamiltonian with eigenvalues corresponding to the observed transition energies and 
\begin{equation}
    H_{int,ij} =  \frac{1}{4\pi \epsilon_0 \epsilon_r}\big[ \frac{\vec{\mu_i} \cdot \vec{\mu_j}}{R_{ij}^3 } - \frac{3(\vec{\mu_i}\cdot\vec{R}_{ij})(\vec{\mu_j}\cdot\vec{R}_{ij})}{R_{ij}^5}\big]
\end{equation}
is the interaction matrix element between pairs of resonant color centers. Here, $\mu$ is taken to be 14.3 Debye \cite{Becker2017}, scaled relative to the maximum dipole moment in the manifold using the linear experimental data for each transition, and oriented randomly along the $\langle111\rangle$ axes of the diamond unit cell; $\vec{R}_{ij} = \vec{r_j} -\vec{r_i}$ is the inter-center separation vector, $R_{ij} = |\vec{R}_{ij}|$ is its magnitude, and $\epsilon_r$ is the relative permittivity of the diamond lattice and is taken to be $\epsilon_r = n^2$ where $n=2.4$ is the index of refraction of diamond at 735 nm \cite{Aggarwal}. The linear response is then calculated by taking the new, shifted peak centers and using those as the center frequency for a Lorentzian lineshape with a full-width, half-maximum corresponding to $T_2 = 120~$ps, the average measured homogeneous electronic dephasing time reported in \cite{Smallwood2020}. Furthermore, we add an interacting, inhomogeneous pedestal corresponding to a Gaussian distribution in center frequency to approximate the measured weak pedestal inhomogeneous linewidth of roughly 1.8~THz \cite{Smallwood2020,Bates2020}. We do this to approximate the influence of a large distribution of dark or weakly luminescent, but interacting centers reported in Ref. \cite{Smallwood2020}. This simulation, presented in Fig.~\ref{fig2} (b)(i), shows that resonant dipole-dipole interactions could contribute to line broadening at first order. The simulation reproduces the main qualitative features of the data and is consistent with the hypothesis that resonant or quasi-resonant dipole-dipole interactions are the main interaction mechanism between adjacent silicon-vacancy centers. 

\begin{figure}
\includegraphics[width = .45\textwidth]{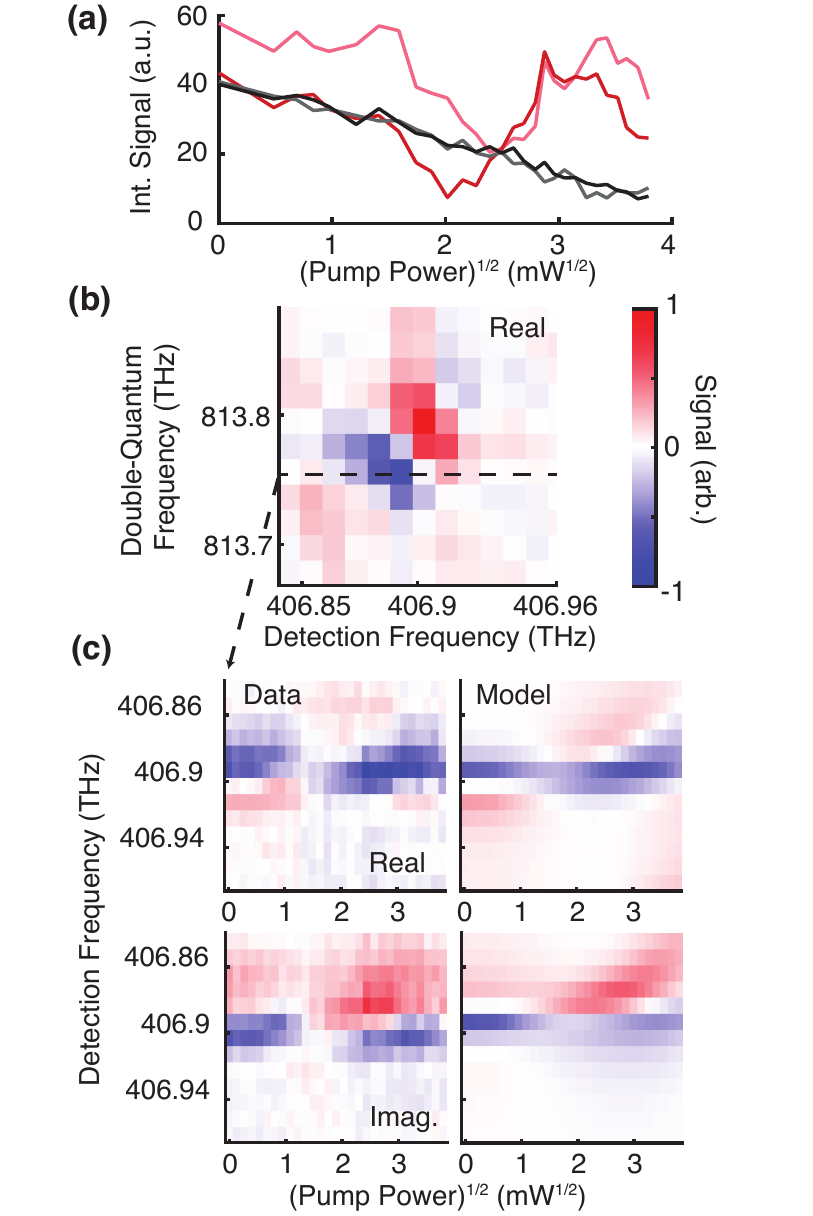}
\caption{(a) The integrated FWM intensity of the peaks color-coded boxes in Fig.~\ref{fig2}(b)(ii) as a function of pump field. (b) The phase-resolved zero-pump upper on-diagonal peak in the red box in Fig.~\ref{fig2}(b)(ii). (c) Slices through the peak in (b) as a function of pump field as compared to a fit with our model.} 
\label{fig3}
\end{figure}

Rudimentary control over inter-center interactions is demonstrated by inducing pairwise Rabi oscillations under the presence of a resonant pump pulse preceding the FWM pulses by 1 ns, allowing coherent artifacts to dephase before the DQ2D probe. We then vary the field of the pump pulse and acquire DQ2D spectra. Figure \ref{fig3}(a) shows the integral of the peaks in the DQ2D spectra as denoted in the boxes in Fig.~\ref{fig2} (b)(ii). The \textit{resonant} center pairs (in the pink and red boxes on the DQ2D diagonal) undergo coherent Rabi oscillations while the non-resonant centers still interact but appear to saturate incoherently as a function of pump field, indicating that the resonant interactions allow pairs of centers to behave as joint, controllable quantum systems.

To show that the oscillations in the signal strength are indeed due to coherent population oscillation in the pairwise four-level systems in our ensemble, we reproduce the qualitative behavior of the lineshape of a one-dimensional slice through the real part of the upper on-diagonal DQ2D peak. Because the nonlinear spectra are inherently phase resolved, we develop a model describing the behavior of this on-diagonal peak corresponding to pairwise, resonant interactions. We choose to fit a slice through the phase-resolved peak in the red box in Fig.~\ref{fig2} (b) (ii), as shown in Fig.~\ref{fig3} (b) and (c). The details of the model are summarized in the supplementary information \cite{supplement}; the most salient detail is that the ground- and excited-state populations of a four-level diamond system scale with driving field as 
\begin{equation}
\rho_{00'} \sim \text{cos}~\Big(\frac{\omega t}{2}\Big)^4 ,\; \rho_{11'} \sim \text{sin}~\Big(\frac{\omega t}{2}\Big)^4. 
\end{equation}

The assumptions of the qualitative model are that the signal is proportional to the ground-state population with interactions scaled linearly in accordance to the excited state population for a given pump field. Explicitly, this corresponds to assuming $S^{(3)}(\tau,\omega_T,\omega_t)\sim \text{cos}^4(\frac{E}{E_\pi})$ and $\Delta = \Delta_{s,0} - i \Delta_{d,0} + \text{sin}^4~(\frac{E}{E_\pi})(\Delta_{s,1} - i \Delta_{d,1})$ where we assume some residual EIS and EID during the DQ2D experiment as quantified in the parameters $\Delta_{s,0}$ and $\Delta_{d,0}$ to account for the zero-pump DQ2D signal. We label $\Delta_{s,1}$ and $ \Delta_{d,1}$ to be the slope of the induced interaction shifts and changes in dephasing respectively. 

We believe that the model is both physically reasonable, in that the interaction parameters obtained from the fit shown in Fig.~\ref{fig3}(c) are within an order of magnitude of the numerical simulations presented in Fig.~\ref{fig2}(b)(i), and that the model reproduces the main qualitative features of the data set. To that end, after geometric, Fresnel, and optical corrections (see the supplement \cite{supplement}) we estimate the pump power required for a pulse area of $\pi$ in our model is $P_\pi=$ 9.6 mW, in agreement with the theoretical value of $P_{\pi,thy}=$ 11.4 mW considering a repetition rate of 75.5 MHz and an estimated FWHM pulse duration of 200 fs. The fitting process returns $\Delta_{d,0} = 2$ GHz, $\Delta_{s,0} = 6$GHz, $\Delta_{d,1} = 50$ GHz, and $\Delta_{s,1} = -200$ GHz, in agreement with an estimation of $\Delta_{s,0} \approx \text{O}(1-10)$ GHz retrieved from our simulations of the linear spectra. 

Importantly, the model reproduces the fact that the peak shape and phase oscillate along with the amplitude reduction and recovery of the signal as the pump field is varied. A more exact model would take into account averaging from the fact that the environment for the pairs of two level systems in our ensemble is likely to be quite inhomogeneous due to the high implantation density. Further, it would account for the fact that the red-boxed peak is composed of two nearly-degenerate peaks from two orientations in the plane of the sample. Such a detailed model would likely quantitatively capture the behavior of the coherent oscillations, but is beyond the scope of this work.

The data presented in this Letter demonstrates that distinct SiV$^-$ centers in ensembles interact. These interactions contribute at a dominant order in the overall optical response of the ensemble and are likely the consequence of dipole-dipole coupling between adjacent centers. Furthermore, quasi-resonant pairs of centers prepared in the same state by a pump pulse can undergo coherent Rabi oscillations as a joint, pairwise system. Similar phenomena have been seen in Rydberg atoms \cite{dudin2012}, but this work constitutes an important demonstration of coherent control of collective systems of multiple color centers. This work opens the way for exploring the utilization of interactions between color centers in proximity. Because they can be deterministically implanted \cite{McLellan2016}, it is perhaps possible to construct lattices of color centers as more experimentally accessible alternatives to trapped ions as test-beds for quantum computation, simulation, or sensing applications.

\begin{acknowledgments}
We thank A. Liu and E. W. Martin for helpful discussions. T. S. acknowledges support from the Federal Ministry of Education and Research of Germany (BMBF, project DiNOQuant, 13N14921). Ion implantation work to generate the SiV– centers was performed, in part, at the Center for Integrated Nanotechnologies, an Office of Science User Facility operated for the U.S. Department of En- ergy (DOE) Office of Science. Sandia National Laboratories is a multi-mission laboratory managed and operated by National Technology and Engineering Solutions of Sandia, LLC., a wholly owned subsidiary of Honeywell International, Inc., for the U.S. Department of Energy’s National Nuclear Security Administration under contract DE-NA-0003525. This paper describes objective technical results and analysis. Any subjective views or opinions that might be expressed in the paper do not necessarily represent the views of the U.S. Department of Energy or the United States Government.
\end{acknowledgments}

%

\bibliography{2Qlib}

\begin{thebibliography}{28}%
\makeatletter
\providecommand \@ifxundefined [1]{%
 \@ifx{#1\undefined}
}%
\providecommand \@ifnum [1]{%
 \ifnum #1\expandafter \@firstoftwo
 \else \expandafter \@secondoftwo
 \fi
}%
\providecommand \@ifx [1]{%
 \ifx #1\expandafter \@firstoftwo
 \else \expandafter \@secondoftwo
 \fi
}%
\providecommand \natexlab [1]{#1}%
\providecommand \enquote  [1]{``#1''}%
\providecommand \bibnamefont  [1]{#1}%
\providecommand \bibfnamefont [1]{#1}%
\providecommand \citenamefont [1]{#1}%
\providecommand \href@noop [0]{\@secondoftwo}%
\providecommand \href [0]{\begingroup \@sanitize@url \@href}%
\providecommand \@href[1]{\@@startlink{#1}\@@href}%
\providecommand \@@href[1]{\endgroup#1\@@endlink}%
\providecommand \@sanitize@url [0]{\catcode `\\12\catcode `\$12\catcode
  `\&12\catcode `\#12\catcode `\^12\catcode `\_12\catcode `\%12\relax}%
\providecommand \@@startlink[1]{}%
\providecommand \@@endlink[0]{}%
\providecommand \url  [0]{\begingroup\@sanitize@url \@url }%
\providecommand \@url [1]{\endgroup\@href {#1}{\urlprefix }}%
\providecommand \urlprefix  [0]{URL }%
\providecommand \Eprint [0]{\href }%
\providecommand \doibase [0]{https://doi.org/}%
\providecommand \selectlanguage [0]{\@gobble}%
\providecommand \bibinfo  [0]{\@secondoftwo}%
\providecommand \bibfield  [0]{\@secondoftwo}%
\providecommand \translation [1]{[#1]}%
\providecommand \BibitemOpen [0]{}%
\providecommand \bibitemStop [0]{}%
\providecommand \bibitemNoStop [0]{.\EOS\space}%
\providecommand \EOS [0]{\spacefactor3000\relax}%
\providecommand \BibitemShut  [1]{\csname bibitem#1\endcsname}%
\let\auto@bib@innerbib\@empty
\bibitem [{\citenamefont {Degen}\ \emph {et~al.}(2017)\citenamefont {Degen},
  \citenamefont {Reinhard},\ and\ \citenamefont {Cappellaro}}]{Degen2017}%
  \BibitemOpen
  \bibfield  {author} {\bibinfo {author} {\bibfnamefont {C.~L.}\ \bibnamefont
  {Degen}}, \bibinfo {author} {\bibfnamefont {F.}~\bibnamefont {Reinhard}},\
  and\ \bibinfo {author} {\bibfnamefont {P.}~\bibnamefont {Cappellaro}},\
  }\bibfield  {title} {\bibinfo {title} {{Quantum sensing}},\ }\href
  {https://doi.org/10.1103/RevModPhys.89.035002} {\bibfield  {journal}
  {\bibinfo  {journal} {Rev. Mod. Phys.}\ }\textbf {\bibinfo {volume} {89}},\
  \bibinfo {pages} {1} (\bibinfo {year} {2017})}\BibitemShut {NoStop}%
\bibitem [{\citenamefont {Sushkov}\ \emph {et~al.}(2014)\citenamefont
  {Sushkov}, \citenamefont {Chisholm}, \citenamefont {Lovchinsky},
  \citenamefont {Kubo}, \citenamefont {Lo}, \citenamefont {Bennett},
  \citenamefont {Hunger}, \citenamefont {Akimov}, \citenamefont {Walsworth},
  \citenamefont {Park},\ and\ \citenamefont {Lukin}}]{Sushkov2014}%
  \BibitemOpen
  \bibfield  {author} {\bibinfo {author} {\bibfnamefont {A.~O.}\ \bibnamefont
  {Sushkov}}, \bibinfo {author} {\bibfnamefont {N.}~\bibnamefont {Chisholm}},
  \bibinfo {author} {\bibfnamefont {I.}~\bibnamefont {Lovchinsky}}, \bibinfo
  {author} {\bibfnamefont {M.}~\bibnamefont {Kubo}}, \bibinfo {author}
  {\bibfnamefont {P.~K.}\ \bibnamefont {Lo}}, \bibinfo {author} {\bibfnamefont
  {S.~D.}\ \bibnamefont {Bennett}}, \bibinfo {author} {\bibfnamefont
  {D.}~\bibnamefont {Hunger}}, \bibinfo {author} {\bibfnamefont
  {A.}~\bibnamefont {Akimov}}, \bibinfo {author} {\bibfnamefont {R.~L.}\
  \bibnamefont {Walsworth}}, \bibinfo {author} {\bibfnamefont {H.}~\bibnamefont
  {Park}},\ and\ \bibinfo {author} {\bibfnamefont {M.~D.}\ \bibnamefont
  {Lukin}},\ }\bibfield  {title} {\bibinfo {title} {{All-optical sensing of a
  single-molecule electron spin}},\ }\href {https://doi.org/10.1021/nl502988n}
  {\bibfield  {journal} {\bibinfo  {journal} {Nano Lett.}\ }\textbf {\bibinfo
  {volume} {14}},\ \bibinfo {pages} {6443} (\bibinfo {year}
  {2014})}\BibitemShut {NoStop}%
\bibitem [{\citenamefont {Nguyen}\ \emph {et~al.}(2018)\citenamefont {Nguyen},
  \citenamefont {Evans}, \citenamefont {Sipahigil}, \citenamefont {Bhaskar},
  \citenamefont {Sukachev}, \citenamefont {Agafonov}, \citenamefont {Davydov},
  \citenamefont {Kulikova}, \citenamefont {Jelezko},\ and\ \citenamefont
  {Lukin}}]{Nguyen2018}%
  \BibitemOpen
  \bibfield  {author} {\bibinfo {author} {\bibfnamefont {C.~T.}\ \bibnamefont
  {Nguyen}}, \bibinfo {author} {\bibfnamefont {R.~E.}\ \bibnamefont {Evans}},
  \bibinfo {author} {\bibfnamefont {A.}~\bibnamefont {Sipahigil}}, \bibinfo
  {author} {\bibfnamefont {M.~K.}\ \bibnamefont {Bhaskar}}, \bibinfo {author}
  {\bibfnamefont {D.~D.}\ \bibnamefont {Sukachev}}, \bibinfo {author}
  {\bibfnamefont {V.~N.}\ \bibnamefont {Agafonov}}, \bibinfo {author}
  {\bibfnamefont {V.~A.}\ \bibnamefont {Davydov}}, \bibinfo {author}
  {\bibfnamefont {L.~F.}\ \bibnamefont {Kulikova}}, \bibinfo {author}
  {\bibfnamefont {F.}~\bibnamefont {Jelezko}},\ and\ \bibinfo {author}
  {\bibfnamefont {M.~D.}\ \bibnamefont {Lukin}},\ }\bibfield  {title} {\bibinfo
  {title} {{All-optical nanoscale thermometry with silicon-vacancy centers in
  diamond}},\ }\href {http://dx.doi.org/10.1063/1.5029904} {\bibfield
  {journal} {\bibinfo  {journal} {Appl. Phys. Lett.}\ }\textbf {\bibinfo
  {volume} {112}} (\bibinfo {year} {2018})}\BibitemShut {NoStop}%
\bibitem [{\citenamefont {Sipahigil}\ \emph {et~al.}(2016)\citenamefont
  {Sipahigil}, \citenamefont {Evans}, \citenamefont {Sukachev}, \citenamefont
  {Burek}, \citenamefont {Borregaard}, \citenamefont {Bhaskar}, \citenamefont
  {Nguyen}, \citenamefont {Pacheco}, \citenamefont {Atikian}, \citenamefont
  {Meuwly}, \citenamefont {Camacho}, \citenamefont {Jelezko}, \citenamefont
  {Bielejec}, \citenamefont {Park}, \citenamefont {Lon{\v{c}}ar},\ and\
  \citenamefont {Lukin}}]{Sipahigil2016}%
  \BibitemOpen
  \bibfield  {author} {\bibinfo {author} {\bibfnamefont {A.}~\bibnamefont
  {Sipahigil}}, \bibinfo {author} {\bibfnamefont {R.~E.}\ \bibnamefont
  {Evans}}, \bibinfo {author} {\bibfnamefont {D.~D.}\ \bibnamefont {Sukachev}},
  \bibinfo {author} {\bibfnamefont {M.~J.}\ \bibnamefont {Burek}}, \bibinfo
  {author} {\bibfnamefont {J.}~\bibnamefont {Borregaard}}, \bibinfo {author}
  {\bibfnamefont {M.~K.}\ \bibnamefont {Bhaskar}}, \bibinfo {author}
  {\bibfnamefont {C.~T.}\ \bibnamefont {Nguyen}}, \bibinfo {author}
  {\bibfnamefont {J.~L.}\ \bibnamefont {Pacheco}}, \bibinfo {author}
  {\bibfnamefont {H.~A.}\ \bibnamefont {Atikian}}, \bibinfo {author}
  {\bibfnamefont {C.}~\bibnamefont {Meuwly}}, \bibinfo {author} {\bibfnamefont
  {R.~M.}\ \bibnamefont {Camacho}}, \bibinfo {author} {\bibfnamefont
  {F.}~\bibnamefont {Jelezko}}, \bibinfo {author} {\bibfnamefont
  {E.}~\bibnamefont {Bielejec}}, \bibinfo {author} {\bibfnamefont
  {H.}~\bibnamefont {Park}}, \bibinfo {author} {\bibfnamefont {M.}~\bibnamefont
  {Lon{\v{c}}ar}},\ and\ \bibinfo {author} {\bibfnamefont {M.~D.}\ \bibnamefont
  {Lukin}},\ }\bibfield  {title} {\bibinfo {title} {{An integrated diamond
  nanophotonics platform for quantum-optical networks}},\ }\href
  {https://doi.org/10.1126/science.aah6875} {\bibfield  {journal} {\bibinfo
  {journal} {Science}\ }\textbf {\bibinfo {volume} {354}},\ \bibinfo {pages}
  {847} (\bibinfo {year} {2016})}\BibitemShut {NoStop}%
\bibitem [{\citenamefont {Sukachev}\ \emph {et~al.}(2017)\citenamefont
  {Sukachev}, \citenamefont {Sipahigil}, \citenamefont {Nguyen}, \citenamefont
  {Bhaskar}, \citenamefont {Evans}, \citenamefont {Jelezko},\ and\
  \citenamefont {Lukin}}]{Sukachev2017}%
  \BibitemOpen
  \bibfield  {author} {\bibinfo {author} {\bibfnamefont {D.~D.}\ \bibnamefont
  {Sukachev}}, \bibinfo {author} {\bibfnamefont {A.}~\bibnamefont {Sipahigil}},
  \bibinfo {author} {\bibfnamefont {C.~T.}\ \bibnamefont {Nguyen}}, \bibinfo
  {author} {\bibfnamefont {M.~K.}\ \bibnamefont {Bhaskar}}, \bibinfo {author}
  {\bibfnamefont {R.~E.}\ \bibnamefont {Evans}}, \bibinfo {author}
  {\bibfnamefont {F.}~\bibnamefont {Jelezko}},\ and\ \bibinfo {author}
  {\bibfnamefont {M.~D.}\ \bibnamefont {Lukin}},\ }\bibfield  {title} {\bibinfo
  {title} {{Silicon-Vacancy Spin Qubit in Diamond: A Quantum Memory Exceeding
  10 ms with Single-Shot State Readout}},\ }\href
  {https://doi.org/10.1103/PhysRevLett.119.223602} {\bibfield  {journal}
  {\bibinfo  {journal} {Phys. Rev. Lett.}\ }\textbf {\bibinfo {volume} {119}},\
  \bibinfo {pages} {1} (\bibinfo {year} {2017})}\BibitemShut {NoStop}%
\bibitem [{\citenamefont {Wan}\ \emph {et~al.}(2020)\citenamefont {Wan},
  \citenamefont {Lu}, \citenamefont {Chen}, \citenamefont {Walsh},
  \citenamefont {Trusheim}, \citenamefont {{De Santis}}, \citenamefont
  {Bersin}, \citenamefont {Harris}, \citenamefont {Mouradian}, \citenamefont
  {Christen}, \citenamefont {Bielejec},\ and\ \citenamefont
  {Englund}}]{Wan2020}%
  \BibitemOpen
  \bibfield  {author} {\bibinfo {author} {\bibfnamefont {N.~H.}\ \bibnamefont
  {Wan}}, \bibinfo {author} {\bibfnamefont {T.~J.}\ \bibnamefont {Lu}},
  \bibinfo {author} {\bibfnamefont {K.~C.}\ \bibnamefont {Chen}}, \bibinfo
  {author} {\bibfnamefont {M.~P.}\ \bibnamefont {Walsh}}, \bibinfo {author}
  {\bibfnamefont {M.~E.}\ \bibnamefont {Trusheim}}, \bibinfo {author}
  {\bibfnamefont {L.}~\bibnamefont {{De Santis}}}, \bibinfo {author}
  {\bibfnamefont {E.~A.}\ \bibnamefont {Bersin}}, \bibinfo {author}
  {\bibfnamefont {I.~B.}\ \bibnamefont {Harris}}, \bibinfo {author}
  {\bibfnamefont {S.~L.}\ \bibnamefont {Mouradian}}, \bibinfo {author}
  {\bibfnamefont {I.~R.}\ \bibnamefont {Christen}}, \bibinfo {author}
  {\bibfnamefont {E.~S.}\ \bibnamefont {Bielejec}},\ and\ \bibinfo {author}
  {\bibfnamefont {D.}~\bibnamefont {Englund}},\ }\bibfield  {title} {\bibinfo
  {title} {{Large-scale integration of artificial atoms in hybrid photonic
  circuits}},\ }\href {https://doi.org/10.1038/s41586-020-2441-3} {\bibfield
  {journal} {\bibinfo  {journal} {Nature}\ }\textbf {\bibinfo {volume} {583}},\
  \bibinfo {pages} {226} (\bibinfo {year} {2020})}\BibitemShut {NoStop}%
\bibitem [{\citenamefont {Bhaskar}\ \emph {et~al.}(2020)\citenamefont
  {Bhaskar}, \citenamefont {Riedinger}, \citenamefont {Machielse},
  \citenamefont {Levonian}, \citenamefont {Nguyen}, \citenamefont {Knall},
  \citenamefont {Park}, \citenamefont {Englund}, \citenamefont {Lon{\v{c}}ar},
  \citenamefont {Sukachev},\ and\ \citenamefont {Lukin}}]{Bhaskar2020}%
  \BibitemOpen
  \bibfield  {author} {\bibinfo {author} {\bibfnamefont {M.~K.}\ \bibnamefont
  {Bhaskar}}, \bibinfo {author} {\bibfnamefont {R.}~\bibnamefont {Riedinger}},
  \bibinfo {author} {\bibfnamefont {B.}~\bibnamefont {Machielse}}, \bibinfo
  {author} {\bibfnamefont {D.~S.}\ \bibnamefont {Levonian}}, \bibinfo {author}
  {\bibfnamefont {C.~T.}\ \bibnamefont {Nguyen}}, \bibinfo {author}
  {\bibfnamefont {E.~N.}\ \bibnamefont {Knall}}, \bibinfo {author}
  {\bibfnamefont {H.}~\bibnamefont {Park}}, \bibinfo {author} {\bibfnamefont
  {D.}~\bibnamefont {Englund}}, \bibinfo {author} {\bibfnamefont
  {M.}~\bibnamefont {Lon{\v{c}}ar}}, \bibinfo {author} {\bibfnamefont {D.~D.}\
  \bibnamefont {Sukachev}},\ and\ \bibinfo {author} {\bibfnamefont {M.~D.}\
  \bibnamefont {Lukin}},\ }\bibfield  {title} {\bibinfo {title} {{Experimental
  demonstration of memory-enhanced quantum communication}},\ }\href
  {https://doi.org/10.1038/s41586-020-2103-5} {\bibfield  {journal} {\bibinfo
  {journal} {Nature}\ }\textbf {\bibinfo {volume} {580}},\ \bibinfo {pages}
  {60} (\bibinfo {year} {2020})}\BibitemShut {NoStop}%
\bibitem [{\citenamefont {Dietrich}\ \emph {et~al.}(2014)\citenamefont
  {Dietrich}, \citenamefont {Jahnke}, \citenamefont {Binder}, \citenamefont
  {Teraji}, \citenamefont {Isoya}, \citenamefont {Rogers},\ and\ \citenamefont
  {Jelezko}}]{Dietrich2014}%
  \BibitemOpen
  \bibfield  {author} {\bibinfo {author} {\bibfnamefont {A.}~\bibnamefont
  {Dietrich}}, \bibinfo {author} {\bibfnamefont {K.~D.}\ \bibnamefont
  {Jahnke}}, \bibinfo {author} {\bibfnamefont {J.~M.}\ \bibnamefont {Binder}},
  \bibinfo {author} {\bibfnamefont {T.}~\bibnamefont {Teraji}}, \bibinfo
  {author} {\bibfnamefont {J.}~\bibnamefont {Isoya}}, \bibinfo {author}
  {\bibfnamefont {L.~J.}\ \bibnamefont {Rogers}},\ and\ \bibinfo {author}
  {\bibfnamefont {F.}~\bibnamefont {Jelezko}},\ }\bibfield  {title} {\bibinfo
  {title} {{Isotopically varying spectral features of silicon-vacancy in
  diamond}},\ }\href@noop {} {\bibfield  {journal} {\bibinfo  {journal} {New J.
  Phys.}\ }\textbf {\bibinfo {volume} {16}} (\bibinfo {year}
  {2014})}\BibitemShut {NoStop}%
\bibitem [{\citenamefont {Liu}\ and\ \citenamefont {Cundiff}(2020)}]{Liu2020}%
  \BibitemOpen
  \bibfield  {author} {\bibinfo {author} {\bibfnamefont {A.}~\bibnamefont
  {Liu}}\ and\ \bibinfo {author} {\bibfnamefont {S.~T.}\ \bibnamefont
  {Cundiff}},\ }\bibfield  {title} {\bibinfo {title} {{Spectroscopic signatures
  of electron-phonon coupling in silicon-vacancy centers in diamond}},\ }\href
  {https://doi.org/10.1103/PhysRevMaterials.4.055202} {\bibfield  {journal}
  {\bibinfo  {journal} {Phys. Rev. Mat.}\ }\textbf {\bibinfo {volume} {4}},\
  \bibinfo {pages} {55202} (\bibinfo {year} {2020})}\BibitemShut {NoStop}%
\bibitem [{\citenamefont {Londero}\ \emph {et~al.}(2018)\citenamefont
  {Londero}, \citenamefont {Thiering}, \citenamefont {Razinkovas},
  \citenamefont {Gali},\ and\ \citenamefont {Alkauskas}}]{Londero2018}%
  \BibitemOpen
  \bibfield  {author} {\bibinfo {author} {\bibfnamefont {E.}~\bibnamefont
  {Londero}}, \bibinfo {author} {\bibfnamefont {G.}~\bibnamefont {Thiering}},
  \bibinfo {author} {\bibfnamefont {L.}~\bibnamefont {Razinkovas}}, \bibinfo
  {author} {\bibfnamefont {A.}~\bibnamefont {Gali}},\ and\ \bibinfo {author}
  {\bibfnamefont {A.}~\bibnamefont {Alkauskas}},\ }\bibfield  {title} {\bibinfo
  {title} {{Vibrational modes of negatively charged silicon-vacancy centers in
  diamond from ab initio calculations}},\ }\href
  {https://doi.org/10.1103/PhysRevB.98.035306} {\bibfield  {journal} {\bibinfo
  {journal} {Phys. Rev. B}\ }\textbf {\bibinfo {volume} {98}},\ \bibinfo
  {pages} {1} (\bibinfo {year} {2018})}\BibitemShut {NoStop}%
\bibitem [{\citenamefont {Meesala}\ \emph {et~al.}(2018)\citenamefont
  {Meesala}, \citenamefont {Sohn}, \citenamefont {Pingault}, \citenamefont
  {Shao}, \citenamefont {Atikian}, \citenamefont {Holzgrafe}, \citenamefont
  {G{\"{u}}ndoğan}, \citenamefont {Stavrakas}, \citenamefont {Sipahigil},
  \citenamefont {Chia}, \citenamefont {Evans}, \citenamefont {Burek},
  \citenamefont {Zhang}, \citenamefont {Wu}, \citenamefont {Pacheco},
  \citenamefont {Abraham}, \citenamefont {Bielejec}, \citenamefont {Lukin},
  \citenamefont {Atat{\"{u}}re},\ and\ \citenamefont {Loncar}}]{Meesala2018}%
  \BibitemOpen
  \bibfield  {author} {\bibinfo {author} {\bibfnamefont {S.}~\bibnamefont
  {Meesala}}, \bibinfo {author} {\bibfnamefont {Y.-I.}\ \bibnamefont {Sohn}},
  \bibinfo {author} {\bibfnamefont {B.}~\bibnamefont {Pingault}}, \bibinfo
  {author} {\bibfnamefont {L.}~\bibnamefont {Shao}}, \bibinfo {author}
  {\bibfnamefont {H.~A.}\ \bibnamefont {Atikian}}, \bibinfo {author}
  {\bibfnamefont {J.}~\bibnamefont {Holzgrafe}}, \bibinfo {author}
  {\bibfnamefont {M.}~\bibnamefont {G{\"{u}}ndoğan}}, \bibinfo {author}
  {\bibfnamefont {C.}~\bibnamefont {Stavrakas}}, \bibinfo {author}
  {\bibfnamefont {A.}~\bibnamefont {Sipahigil}}, \bibinfo {author}
  {\bibfnamefont {C.}~\bibnamefont {Chia}}, \bibinfo {author} {\bibfnamefont
  {R.}~\bibnamefont {Evans}}, \bibinfo {author} {\bibfnamefont {M.~J.}\
  \bibnamefont {Burek}}, \bibinfo {author} {\bibfnamefont {M.}~\bibnamefont
  {Zhang}}, \bibinfo {author} {\bibfnamefont {L.}~\bibnamefont {Wu}}, \bibinfo
  {author} {\bibfnamefont {J.~L.}\ \bibnamefont {Pacheco}}, \bibinfo {author}
  {\bibfnamefont {J.}~\bibnamefont {Abraham}}, \bibinfo {author} {\bibfnamefont
  {E.}~\bibnamefont {Bielejec}}, \bibinfo {author} {\bibfnamefont {M.~D.}\
  \bibnamefont {Lukin}}, \bibinfo {author} {\bibfnamefont {M.}~\bibnamefont
  {Atat{\"{u}}re}},\ and\ \bibinfo {author} {\bibfnamefont {M.}~\bibnamefont
  {Loncar}},\ }\bibfield  {title} {\bibinfo {title} {{Strain engineering of the
  silicon-vacancy center in diamond}},\ }\href
  {https://journals.aps.org/prb/pdf/10.1103/PhysRevB.97.205444} {\bibfield
  {journal} {\bibinfo  {journal} {Phys. Rev. B}\ }\textbf {\bibinfo {volume}
  {97}} (\bibinfo {year} {2018})}\BibitemShut {NoStop}%
\bibitem [{\citenamefont {Udvarhelyi}\ \emph {et~al.}(2019)\citenamefont
  {Udvarhelyi}, \citenamefont {Nagy}, \citenamefont {Kaiser}, \citenamefont
  {Lee}, \citenamefont {Wrachtrup},\ and\ \citenamefont
  {Gali}}]{Udvarhelyi2019}%
  \BibitemOpen
  \bibfield  {author} {\bibinfo {author} {\bibfnamefont {P.}~\bibnamefont
  {Udvarhelyi}}, \bibinfo {author} {\bibfnamefont {R.}~\bibnamefont {Nagy}},
  \bibinfo {author} {\bibfnamefont {F.}~\bibnamefont {Kaiser}}, \bibinfo
  {author} {\bibfnamefont {S.-Y.}\ \bibnamefont {Lee}}, \bibinfo {author}
  {\bibfnamefont {J.}~\bibnamefont {Wrachtrup}},\ and\ \bibinfo {author}
  {\bibfnamefont {A.}~\bibnamefont {Gali}},\ }\bibfield  {title} {\bibinfo
  {title} {Spectrally stable defect qubits with no inversion symmetry for
  robust spin-to-photon interface},\ }\href
  {https://doi.org/10.1103/PhysRevApplied.11.044022} {\bibfield  {journal}
  {\bibinfo  {journal} {Phys. Rev. Appl.}\ }\textbf {\bibinfo {volume} {11}},\
  \bibinfo {pages} {044022} (\bibinfo {year} {2019})}\BibitemShut {NoStop}%
\bibitem [{\citenamefont {Campbell}\ and\ \citenamefont
  {Hudson}(2020)}]{campbell2020}%
  \BibitemOpen
  \bibfield  {author} {\bibinfo {author} {\bibfnamefont {W.~C.}\ \bibnamefont
  {Campbell}}\ and\ \bibinfo {author} {\bibfnamefont {E.~R.}\ \bibnamefont
  {Hudson}},\ }\bibfield  {title} {\bibinfo {title} {{Dipole-Phonon Quantum
  Logic with Trapped Polar Molecular Ions}},\ }\href
  {https://doi.org/10.1103/PhysRevLett.125.120501} {\bibfield  {journal}
  {\bibinfo  {journal} {Phys. Rev. Lett.}\ }\textbf {\bibinfo {volume} {125}},\
  \bibinfo {pages} {120501} (\bibinfo {year} {2020})}\BibitemShut {NoStop}%
\bibitem [{\citenamefont {Evans}\ \emph {et~al.}(2018)\citenamefont {Evans},
  \citenamefont {Bhaskar}, \citenamefont {Sukachev}, \citenamefont {Nguyen},
  \citenamefont {Sipahigil}, \citenamefont {Burek}, \citenamefont {Machielse},
  \citenamefont {Zhang}, \citenamefont {Zibrov}, \citenamefont {Bielejec},
  \citenamefont {Park}, \citenamefont {Lon{\v{c}}ar},\ and\ \citenamefont
  {Lukin}}]{Evans2018}%
  \BibitemOpen
  \bibfield  {author} {\bibinfo {author} {\bibfnamefont {R.~E.}\ \bibnamefont
  {Evans}}, \bibinfo {author} {\bibfnamefont {M.~K.}\ \bibnamefont {Bhaskar}},
  \bibinfo {author} {\bibfnamefont {D.~D.}\ \bibnamefont {Sukachev}}, \bibinfo
  {author} {\bibfnamefont {C.~T.}\ \bibnamefont {Nguyen}}, \bibinfo {author}
  {\bibfnamefont {A.}~\bibnamefont {Sipahigil}}, \bibinfo {author}
  {\bibfnamefont {M.~J.}\ \bibnamefont {Burek}}, \bibinfo {author}
  {\bibfnamefont {B.}~\bibnamefont {Machielse}}, \bibinfo {author}
  {\bibfnamefont {G.~H.}\ \bibnamefont {Zhang}}, \bibinfo {author}
  {\bibfnamefont {A.~S.}\ \bibnamefont {Zibrov}}, \bibinfo {author}
  {\bibfnamefont {E.}~\bibnamefont {Bielejec}}, \bibinfo {author}
  {\bibfnamefont {H.}~\bibnamefont {Park}}, \bibinfo {author} {\bibfnamefont
  {M.}~\bibnamefont {Lon{\v{c}}ar}},\ and\ \bibinfo {author} {\bibfnamefont
  {M.~D.}\ \bibnamefont {Lukin}},\ }\bibfield  {title} {\bibinfo {title}
  {{Photon-mediated interactions between quantum emitters in a diamond
  nanocavity}},\ }\href {https://doi.org/10.1126/science.aau4691} {\bibfield
  {journal} {\bibinfo  {journal} {Science}\ }\textbf {\bibinfo {volume}
  {362}},\ \bibinfo {pages} {662} (\bibinfo {year} {2018})}\BibitemShut
  {NoStop}%
\bibitem [{\citenamefont {Kim}\ \emph {et~al.}(2009)\citenamefont {Kim},
  \citenamefont {Mukamel},\ and\ \citenamefont {Scholes}}]{Kim2009}%
  \BibitemOpen
  \bibfield  {author} {\bibinfo {author} {\bibfnamefont {J.}~\bibnamefont
  {Kim}}, \bibinfo {author} {\bibfnamefont {S.}~\bibnamefont {Mukamel}},\ and\
  \bibinfo {author} {\bibfnamefont {G.~D.}\ \bibnamefont {Scholes}},\
  }\bibfield  {title} {\bibinfo {title} {{Two-dimensional electronic
  double-quantum coherence spectroscopy.}},\ }\href
  {https://doi.org/10.1021/ar9000795} {\bibfield  {journal} {\bibinfo
  {journal} {Accounts of chemical research}\ }\textbf {\bibinfo {volume}
  {42}},\ \bibinfo {pages} {1375} (\bibinfo {year} {2009})}\BibitemShut
  {NoStop}%
\bibitem [{\citenamefont {Bates}\ \emph {et~al.}(2021)\citenamefont {Bates},
  \citenamefont {Day}, \citenamefont {Smallwood}, \citenamefont {Owen},
  \citenamefont {Ulbrichcht}, \citenamefont {Schr\"oder}, \citenamefont
  {Autry}, \citenamefont {Diederich}, \citenamefont {Bielejec}, \citenamefont
  {Siemens},\ and\ \citenamefont {Cundiff}}]{Bates2020}%
  \BibitemOpen
  \bibfield  {author} {\bibinfo {author} {\bibfnamefont {K.~M.}\ \bibnamefont
  {Bates}}, \bibinfo {author} {\bibfnamefont {M.~W.}\ \bibnamefont {Day}},
  \bibinfo {author} {\bibfnamefont {C.~L.}\ \bibnamefont {Smallwood}}, \bibinfo
  {author} {\bibfnamefont {R.~C.}\ \bibnamefont {Owen}}, \bibinfo {author}
  {\bibfnamefont {R.}~\bibnamefont {Ulbrichcht}}, \bibinfo {author}
  {\bibfnamefont {T.}~\bibnamefont {Schr\"oder}}, \bibinfo {author}
  {\bibfnamefont {T.~M.}\ \bibnamefont {Autry}}, \bibinfo {author}
  {\bibfnamefont {G.}~\bibnamefont {Diederich}}, \bibinfo {author}
  {\bibfnamefont {E.}~\bibnamefont {Bielejec}}, \bibinfo {author}
  {\bibfnamefont {M.~E.}\ \bibnamefont {Siemens}},\ and\ \bibinfo {author}
  {\bibfnamefont {S.~T.}\ \bibnamefont {Cundiff}},\ }\href@noop {} {\
  (\bibinfo {year} {2021})},\ \Eprint {https://arxiv.org/abs/2104.09638}
  {arXiv:2104.09638} \BibitemShut {NoStop}%
\bibitem [{sup()}]{supplement}%
  \BibitemOpen
  \href@noop {} {\bibinfo {title} {See {Supplemental Material} at
  {http://link.aps.org/ supplemental/XXXX} and references within
  \cite{Rogers2014,Smallwood2016}}}\BibitemShut {NoStop}%
\bibitem [{\citenamefont {Smallwood}\ \emph {et~al.}(2016)\citenamefont
  {Smallwood}, \citenamefont {Autry},\ and\ \citenamefont
  {Cundiff}}]{Smallwood2016}%
  \BibitemOpen
  \bibfield  {author} {\bibinfo {author} {\bibfnamefont {C.}~\bibnamefont
  {Smallwood}}, \bibinfo {author} {\bibfnamefont {T.}~\bibnamefont {Autry}},\
  and\ \bibinfo {author} {\bibfnamefont {S.}~\bibnamefont {Cundiff}},\
  }\bibfield  {title} {\bibinfo {title} {{Analytical solutions to the
  finite-pulse Bloch model for multidimensional coherent spectroscopy}},\
  }\href@noop {} {\bibfield  {journal} {\bibinfo  {journal} {J. Opt. Soc. Am.
  B}\ }\textbf {\bibinfo {volume} {34}} (\bibinfo {year} {2016})}\BibitemShut
  {NoStop}%
\bibitem [{\citenamefont {Tekavec}\ \emph {et~al.}(2007)\citenamefont
  {Tekavec}, \citenamefont {Lott},\ and\ \citenamefont {Marcus}}]{Tekavec2007}%
  \BibitemOpen
  \bibfield  {author} {\bibinfo {author} {\bibfnamefont {P.~F.}\ \bibnamefont
  {Tekavec}}, \bibinfo {author} {\bibfnamefont {G.~A.}\ \bibnamefont {Lott}},\
  and\ \bibinfo {author} {\bibfnamefont {A.~H.}\ \bibnamefont {Marcus}},\
  }\bibfield  {title} {\bibinfo {title} {{Fluorescence-detected two-dimensional
  electronic coherence spectroscopy by acousto-optic phase modulation}},\
  }\href {https://doi.org/10.1063/1.2800560} {\bibfield  {journal} {\bibinfo
  {journal} {J. Chem. Phys.}\ }\textbf {\bibinfo {volume} {127}},\ \bibinfo
  {pages} {1} (\bibinfo {year} {2007})}\BibitemShut {NoStop}%
\bibitem [{\citenamefont {Nardin}\ \emph {et~al.}(2013)\citenamefont {Nardin},
  \citenamefont {Autry}, \citenamefont {Silverman},\ and\ \citenamefont
  {Cundiff}}]{Nardin2013}%
  \BibitemOpen
  \bibfield  {author} {\bibinfo {author} {\bibfnamefont {G.}~\bibnamefont
  {Nardin}}, \bibinfo {author} {\bibfnamefont {T.~M.}\ \bibnamefont {Autry}},
  \bibinfo {author} {\bibfnamefont {K.~L.}\ \bibnamefont {Silverman}},\ and\
  \bibinfo {author} {\bibfnamefont {S.~T.}\ \bibnamefont {Cundiff}},\
  }\bibfield  {title} {\bibinfo {title} {{Multidimensional coherent
  photocurrent spectroscopy of a semiconductor nanostructure.}},\ }\href
  {https://doi.org/10.1364/OE.21.028617} {\bibfield  {journal} {\bibinfo
  {journal} {Opt. Express}\ }\textbf {\bibinfo {volume} {21}},\ \bibinfo
  {pages} {28617} (\bibinfo {year} {2013})}\BibitemShut {NoStop}%
\bibitem [{\citenamefont {Smallwood}\ \emph {et~al.}(2020)\citenamefont
  {Smallwood}, \citenamefont {Ulbricht}, \citenamefont {Day}, \citenamefont
  {Schr{\"{o}}der}, \citenamefont {Bates}, \citenamefont {Autry}, \citenamefont
  {Diederich}, \citenamefont {Bielejec}, \citenamefont {Siemens},\ and\
  \citenamefont {Cundiff}}]{Smallwood2020}%
  \BibitemOpen
  \bibfield  {author} {\bibinfo {author} {\bibfnamefont {C.~L.}\ \bibnamefont
  {Smallwood}}, \bibinfo {author} {\bibfnamefont {R.}~\bibnamefont {Ulbricht}},
  \bibinfo {author} {\bibfnamefont {M.~W.}\ \bibnamefont {Day}}, \bibinfo
  {author} {\bibfnamefont {T.}~\bibnamefont {Schr{\"{o}}der}}, \bibinfo
  {author} {\bibfnamefont {K.~M.}\ \bibnamefont {Bates}}, \bibinfo {author}
  {\bibfnamefont {T.~M.}\ \bibnamefont {Autry}}, \bibinfo {author}
  {\bibfnamefont {G.}~\bibnamefont {Diederich}}, \bibinfo {author}
  {\bibfnamefont {E.}~\bibnamefont {Bielejec}}, \bibinfo {author}
  {\bibfnamefont {M.~E.}\ \bibnamefont {Siemens}},\ and\ \bibinfo {author}
  {\bibfnamefont {S.~T.}\ \bibnamefont {Cundiff}},\ }\bibfield  {title}
  {\bibinfo {title} {{Hidden Silicon-Vacancy Centers in Diamond}},\ }\href
  {http://arxiv.org/abs/2006.02323} {\  (\bibinfo {year} {2020})},\ \Eprint
  {https://arxiv.org/abs/2006.02323} {arXiv:2006.02323} \BibitemShut {NoStop}%
\bibitem [{\citenamefont {Specht}\ \emph {et~al.}(2015)\citenamefont {Specht},
  \citenamefont {Knorr},\ and\ \citenamefont {Richter}}]{specht2015}%
  \BibitemOpen
  \bibfield  {author} {\bibinfo {author} {\bibfnamefont {J.~F.}\ \bibnamefont
  {Specht}}, \bibinfo {author} {\bibfnamefont {A.}~\bibnamefont {Knorr}},\ and\
  \bibinfo {author} {\bibfnamefont {M.}~\bibnamefont {Richter}},\ }\bibfield
  {title} {\bibinfo {title} {{Two-dimensional spectroscopy: An approach to
  distinguish F{\"{o}}rster and Dexter transfer processes in coupled
  nanostructures}},\ }\href {https://doi.org/10.1103/PhysRevB.91.155313}
  {\bibfield  {journal} {\bibinfo  {journal} {Phys. Rev. B}\ }\textbf {\bibinfo
  {volume} {91}},\ \bibinfo {pages} {1} (\bibinfo {year} {2015})}\BibitemShut
  {NoStop}%
\bibitem [{\citenamefont {Nardin}\ \emph {et~al.}(2014)\citenamefont {Nardin},
  \citenamefont {Moody}, \citenamefont {Singh}, \citenamefont {Autry},
  \citenamefont {Li}, \citenamefont {Morier-Genoud},\ and\ \citenamefont
  {Cundiff}}]{Nardin2014}%
  \BibitemOpen
  \bibfield  {author} {\bibinfo {author} {\bibfnamefont {G.}~\bibnamefont
  {Nardin}}, \bibinfo {author} {\bibfnamefont {G.}~\bibnamefont {Moody}},
  \bibinfo {author} {\bibfnamefont {R.}~\bibnamefont {Singh}}, \bibinfo
  {author} {\bibfnamefont {T.~M.}\ \bibnamefont {Autry}}, \bibinfo {author}
  {\bibfnamefont {H.}~\bibnamefont {Li}}, \bibinfo {author} {\bibfnamefont
  {F.}~\bibnamefont {Morier-Genoud}},\ and\ \bibinfo {author} {\bibfnamefont
  {S.~T.}\ \bibnamefont {Cundiff}},\ }\bibfield  {title} {\bibinfo {title}
  {{Coherent excitonic coupling in an asymmetric double InGaAs quantum well
  arises from many-body effects}},\ }\href@noop {} {\bibfield  {journal}
  {\bibinfo  {journal} {Phys. Rev. Lett.}\ }\textbf {\bibinfo {volume} {112}},\
  \bibinfo {pages} {046402} (\bibinfo {year} {2014})}\BibitemShut {NoStop}%
\bibitem [{\citenamefont {Gali}\ and\ \citenamefont {Maze}(2013)}]{Gali2013}%
  \BibitemOpen
  \bibfield  {author} {\bibinfo {author} {\bibfnamefont {A.}~\bibnamefont
  {Gali}}\ and\ \bibinfo {author} {\bibfnamefont {J.~R.}\ \bibnamefont
  {Maze}},\ }\bibfield  {title} {\bibinfo {title} {{Ab initio study of the
  split silicon-vacancy defect in diamond: Electronic structure and related
  properties}},\ }\href@noop {} {\bibfield  {journal} {\bibinfo  {journal}
  {Phys. Rev. B}\ }\textbf {\bibinfo {volume} {88}},\ \bibinfo {pages} {1}
  (\bibinfo {year} {2013})}\BibitemShut {NoStop}%
\bibitem [{\citenamefont {Becker}\ and\ \citenamefont
  {Becher}(2017)}]{Becker2017}%
  \BibitemOpen
  \bibfield  {author} {\bibinfo {author} {\bibfnamefont {J.~N.}\ \bibnamefont
  {Becker}}\ and\ \bibinfo {author} {\bibfnamefont {C.}~\bibnamefont
  {Becher}},\ }\bibfield  {title} {\bibinfo {title} {{Coherence Properties and
  Quantum Control of Silicon Vacancy Color Centers in Diamond}},\ }\href
  {https://doi.org/10.1002/pssa.201700586} {\bibfield  {journal} {\bibinfo
  {journal} {Phys. Status Solidi A}\ }\textbf {\bibinfo {volume} {214}},\
  \bibinfo {pages} {1} (\bibinfo {year} {2017})}\BibitemShut {NoStop}%
\bibitem [{\citenamefont {Aggarwal}\ and\ \citenamefont
  {Ramdas}(2019)}]{Aggarwal}%
  \BibitemOpen
  \bibfield  {author} {\bibinfo {author} {\bibfnamefont {R.~L.}\ \bibnamefont
  {Aggarwal}}\ and\ \bibinfo {author} {\bibfnamefont {A.~K.}\ \bibnamefont
  {Ramdas}},\ }\href@noop {} {\emph {\bibinfo {title} {Physical properties of
  diamond and sapphire}}}\ (\bibinfo  {publisher} {CRC Press/Taylor \& Francis
  Group},\ \bibinfo {address} {Boca Raton, FL},\ \bibinfo {year}
  {2019})\BibitemShut {NoStop}%
\bibitem [{\citenamefont {Dudin}\ \emph {et~al.}(2012)\citenamefont {Dudin},
  \citenamefont {Li}, \citenamefont {Bariani},\ and\ \citenamefont
  {Kuzmich}}]{dudin2012}%
  \BibitemOpen
  \bibfield  {author} {\bibinfo {author} {\bibfnamefont {Y.~O.}\ \bibnamefont
  {Dudin}}, \bibinfo {author} {\bibfnamefont {L.}~\bibnamefont {Li}}, \bibinfo
  {author} {\bibfnamefont {F.}~\bibnamefont {Bariani}},\ and\ \bibinfo {author}
  {\bibfnamefont {A.}~\bibnamefont {Kuzmich}},\ }\bibfield  {title} {\bibinfo
  {title} {{Observation of coherent many-body Rabi oscillations}},\ }\href
  {https://doi.org/10.1038/nphys2413} {\bibfield  {journal} {\bibinfo
  {journal} {Nat. Phys.}\ }\textbf {\bibinfo {volume} {8}},\ \bibinfo {pages}
  {790} (\bibinfo {year} {2012})},\ \Eprint {https://arxiv.org/abs/1205.7061}
  {1205.7061} \BibitemShut {NoStop}%
\bibitem [{\citenamefont {McLellan}\ \emph {et~al.}(2016)\citenamefont
  {McLellan}, \citenamefont {Myers}, \citenamefont {Kraemer}, \citenamefont
  {Ohno}, \citenamefont {Awschalom},\ and\ \citenamefont {{Bleszynski
  Jayich}}}]{McLellan2016}%
  \BibitemOpen
  \bibfield  {author} {\bibinfo {author} {\bibfnamefont {C.~A.}\ \bibnamefont
  {McLellan}}, \bibinfo {author} {\bibfnamefont {B.~A.}\ \bibnamefont {Myers}},
  \bibinfo {author} {\bibfnamefont {S.}~\bibnamefont {Kraemer}}, \bibinfo
  {author} {\bibfnamefont {K.}~\bibnamefont {Ohno}}, \bibinfo {author}
  {\bibfnamefont {D.~D.}\ \bibnamefont {Awschalom}},\ and\ \bibinfo {author}
  {\bibfnamefont {A.~C.}\ \bibnamefont {{Bleszynski Jayich}}},\ }\bibfield
  {title} {\bibinfo {title} {{Patterned Formation of Highly Coherent
  Nitrogen-Vacancy Centers Using a Focused Electron Irradiation Technique}},\
  }\href {https://doi.org/10.1021/acs.nanolett.5b05304} {\bibfield  {journal}
  {\bibinfo  {journal} {Nano Lett.}\ }\textbf {\bibinfo {volume} {16}},\
  \bibinfo {pages} {2450} (\bibinfo {year} {2016})}\BibitemShut {NoStop}%
\end{thebibliography}%

\end{document}